\documentstyle[preprint,tighten,floats,prd,aps]{revtex}
\input epsf.tex
\def\DESepsf(#1 width #2){\epsfxsize=#2 \epsfbox{#1}}
\begin{document}
\draft
\preprint{\vbox{
\hbox{hep-ph/9511258}}  }
\title{\bf On the QCD perturbative expansion for $e^+e^- \to$ hadrons}
\bigskip
\author{ Davison E.\ Soper}
\address{
Institute of Theoretical Science\\
University of Oregon,
Eugene, OR  97403
}
\author{ Levan R.\ Surguladze}
\address{
Department of Physics and Astronomy\\
University of Alabama,
Tuscaloosa, AL 35487
}
\date{7 November 1995}
\maketitle

\begin{abstract}

We study the perturbative QCD series for the hadronic width of the $Z$
boson. We sum a class of large ``$\pi^2$ terms'' and reorganize the
series so as to minimize ``renormalon'' effects. We also consider the
renormalization scheme-scale ambiguity of the perturbative results.
We find that, with three nontrivial known terms in the perturbative
expansion, the treatment of the $\pi^2$ terms is quite
important, while renormalon effects are less important. The measured
hadronic width of the $Z$ is often used to determine the value of
$\alpha_s(M_Z^2)$. A standard method is to use the perturbative
expansion for the width truncated at order $\alpha_s^3$ in the
$\overline{\rm MS}$ scheme with scale $\mu = M_Z$. We estimate that
the determined value of $\alpha_s(M_Z^2)$ should be increased by 0.6\%
compared to the value extracted with this standard method. After this
adjustment for $\pi^2$ and renormalon effects, we estimate that
the uncertainty in $\alpha_s(M_Z^2)$ arising from QCD theory is about
$0.4\%$. This is, of course, much less than the experimental
uncertainty of about 5\%.

\end{abstract}
\pacs{}
\narrowtext
\section{Introduction}
\label{Introduction}

The width for $Z \to {\rm hadrons}$ is conventionally described by
the ratio $R$ of this width to the width for $Z \to e^+ e^-$. The $Z$
boson need not be on-shell: for theoretical purposes, we can consider
$R$ as a function $R(s)$ of the c.m.\ energy $s$ of the $e^+e^-$
annihilation that produces the $Z$. Then the measured $R$ is $R(M_Z^2)$.
One way of measuring the strong coupling $\alpha_s$ is to compare theory
and experiment for $R(M_Z^2)$. The purpose of this paper is to discuss
some aspects of the theoretical evaluation of $R(s)$: the effect of
``$\pi^2$ terms'' and  ``renormalons'' on the determination of $R$ from
the calculated terms in its perturbative expansion in powers of
$\alpha_s$. Our goal is to suggest ways of evaluating $R(s)$ as
precisely as possible from the knowledge of the first three terms in its
perturbative expansion and then to estimate the theoretical error in
this evaluation. We pose the question of whether $\alpha_s(M_Z^2)$ could
be extracted at a precision of a few parts per mill from $R(M_Z^2)$ in
the hypothetical case that infinitely accurate data were available and
uncertainties in the electroweak part of the calculation were zero.

We will conclude that a QCD theoretical error on $\alpha_s(M_Z^2)$ of
about four parts per mill is possible if one understands this as a
one $\sigma$ error estimate: the QCD error is probably about this size.
An estimate of the QCD theoretical error at the 95\% confidence level
would be quite a lot larger because it should include the possibility
that certain hypotheses -- guesses really -- about the behavior of the
perturbative expansion are simply wrong. We will try to make clear the
nature of the required hypotheses and let the reader form his or her own
judgment.

In this paper, we adopt a simplified theoretical framework so that
we can concentrate on the QCD effects. We consider $Z \to {\rm hadrons}$
at the Born level in the electroweak interactions. We take the $u$, $d$,
$s$, $c$ and $b$ quarks to be exactly massless. We include contributions
from virtual top quarks that behave like $\log^n(m_t^2/M_Z^2)$, dropping
terms that behave like $(M_Z^2/m_t^2)^n$ as $m_t \to \infty$.

Given this theoretical framework, the theoretical expression for
$R(M_Z^2)$ has the form
\begin{equation}
R(M_Z^2) = R_0\left\{ 1 + {\cal R}(M_Z^2)\right\}.
\end{equation}
Here $R_0$ is the value of $R$ in the parton model, without perturbative
QCD corrections. The QCD corrections are contained in
${\cal R}(M_Z^2) = \alpha_s(M_Z^2)/\pi + {\cal O}(\alpha_s^2)$, which is
often denoted $\delta_{\rm QCD}$. We study ${\cal R}(M_Z^2)$ and try to
estimate the theoretical uncertainty in  ${\cal R}(M_Z^2)$ caused by
evaluating it in perturbation theory truncated at order $\alpha_s^3$.
For this purpose, we use a nominal value $\overline\alpha_s(M_Z^2) =
0.120$ of the $\overline{\rm MS}$ strong coupling evaluated at $M_Z$. If
an experimental value for ${\cal R}(M_Z^2)$ were used to extract
$\overline\alpha_s(M_Z^2)$, then the fractional theoretical uncertainty
in  ${\cal R}(M_Z^2)$ would translate into a fractional uncertainty of
the same size for $\alpha_s(M_Z^2)$.

When we present numerical results, we choose $M_Z = 91.188 {\ \rm GeV}$
and $\sin^2 \theta_W = 0.2319$. We take the top quark pole mass to be
170 GeV, as estimated in Ref.~\cite{DESMoriond} from the CDF and D0
results \cite{CDFD0}.

The scope of this paper is limited, and in fact we do not attempt to
evaluate ${\cal R}(M_Z^2)$ at the level of precision that we are
discussing. Such an evaluation involves careful consideration of a large
number of small effects. Among these are electroweak effects beyond the
Born level \cite{revZ}, effects of non-zero masses for the light quarks
\cite{O(m)} and $(M_Z^2/m_t^2)^n$ contributions from virtual loops
containing the top quark \cite{Mt}. We review the status of some of
these issues in the appendix to this paper.

\section{The running coupling and top mass}
\label{RunningCoupling}

In this paper we denote by $\alpha_s(s)$ the running coupling in a
renormalization scheme that may or may not be the
$\overline {\rm MS}$ scheme \cite{MS}. We denote by
$\overline\alpha_s(s)$ the running coupling as defined by the
$\overline {\rm MS}$ scheme with five flavors of light quarks.

The dependence of $\alpha_s(s e^t)$ on $t$ is given by the
renormalization group equation
\begin{eqnarray}
{ d \over d t}
\left(\pi \over \alpha_s(se^t)\right)
&=&-{\pi^2\over
\alpha_s^2(se^t)}\ \beta(\alpha_s(se^t))
\nonumber\\
&=&
\beta_0
+\beta_1\ {\alpha_s(se^t) \over \pi}
+\beta_2\left(\alpha_s(se^t) \over \pi\right)^2
+\cdots.
\label{rengroup}
\end{eqnarray}
We use this equation to derive an approximation for ${\pi / \alpha_s(s
e^t)}$. We find
\begin{eqnarray}
{\pi \over \alpha_s(s e^t)}
&=&{\pi \over \alpha_s(s)}\,(1+x)
+ { \beta_1 \over \beta_0}\log(1+x)
\nonumber\\
&&+{ \alpha_s(s) \over \pi}
\left(
{ \beta_0\beta_2 - \beta_1^2 \over \beta_0^2}\
{ x \over 1+x}
+{ \beta_1^2 \over \beta_0^2}\
{ \log(1+x) \over 1+x}
\right)
\nonumber\\
&&+\cdots,
\label{runningalpha}
%
\end{eqnarray}
where
\begin{equation}
x = \beta_0 t\,{ \alpha_s(s) \over \pi}.
\end{equation}
Further terms in this series involve higher powers of
$\alpha_s(s)/\pi$ times functions of $x$ that are proportional to $x$
for small $x$. We do not include any more terms because the next term
involves the coefficient $\beta_3$, which is unknown. If we wanted to
recover the ordinary perturbative expansion of ${\pi / \alpha_s(s
e^t)}$ up to order $\alpha_s(s)^2$, we would note that $x$ is
proportional to $\alpha_s$ and expand in powers of $x$, then omit terms
beyond $x^2$ or $x\alpha_s$. Eq.~(\ref{runningalpha}) is better than the
purely perturbative expansion because it is a valid expansion in powers
of $\alpha_s(s)$ when $x$ is fixed at some finite value. Thus it is
useful when $\beta_0 t$ is as large as $\pi/\alpha_s(s)$.

We shall sometimes want to examine the dependence of the
results of calculations on the renormalization scheme used in the
calculation ({\it Cf.} Ref.~\cite{Stev}). For this purpose, we define
an $\alpha_s$ in a renormalization scheme that may not be the
$\overline {\rm MS}$ scheme by
\begin{equation}
\alpha_s(s) = \overline\alpha_s(s)
+ c_2\, \overline\alpha_s(s)^2
+ c_3\, \overline\alpha_s(s)^3 + \cdots.
\label{alphachange0}
\end{equation}
Then one can use $\alpha_s(s)$ as the expansion parameter of the
theory. Since the perturbative formulas used are inevitably
truncated at some order of perturbation theory, the results depend
on the coefficients $c_i$ that specify the scheme. We will want to
find out how much the results depend on the $c_i$. There are two
purposes to this. First, the choice of renormalization scheme
represents an ambiguity of the theory, and we want to have an
estimate of the numerical importance of this ambiguity. Second,
there are uncalculated higher order terms that are, by necessity,
omitted from the calculation. Parts of these terms serve to cancel
the dependence of the results on the $c_i$. Thus the observed size
of the dependence of the result on the $c_i$ serves as a rough
indicator of the size of the uncalculated higher order terms.

The coefficient $c_2$ can be simply absorbed into a change of the
scale of the running coupling:
\begin{equation}
\alpha_s(s) = \overline\alpha_s(se^{\delta t})
+ c_3^\prime\, \overline\alpha_s^3(se^{\delta t}) + \cdots.
\label{alphachange}
\end{equation}
That is, using Eq.~(\ref{runningalpha}) on the right hand side of
Eq.~(\ref{alphachange}) reproduces Eq.~(\ref{alphachange0}).
The term in Eq.~(\ref{alphachange}) proportional to
$\overline\alpha_s^3$ results in a change of the coefficient $\beta_2$
in the $\beta$ function that describes the running of $\alpha_s$.
(Recall that $\beta_0$ and $\beta_1$ are scheme independent.) Let us
parameterize this change as
\begin{equation}
\beta_2 = \overline\beta_2 + \delta\beta_2\,,
\end{equation}
where  $\overline\beta_2$ is the third coefficient of the $\beta$
function in the $\overline{\rm MS}$ scheme and other MS-type schemes.
Then the relation between $\alpha_s$ and $\overline\alpha_s$ can be
written as
\begin{equation}
{\alpha_s(s)\over \pi} =
{\overline\alpha_s(se^{\delta t}) \over \pi}
+ { \delta\beta_2 \over \beta_0}\,
\left (\overline\alpha_s(se^{\delta t})\over \pi\right)^3 +
\cdots.
\label{alphasdef}
\end{equation}
We shall use $\delta t$ and $\delta\beta_2$ to parameterize the
choice of scheme.

By combining Eq.~(\ref{runningalpha}) with Eq.~(\ref{alphasdef}),  we see
that $\alpha_s(s)$ can be expanded in terms of $\overline\alpha_s(s)$
by using
\begin{eqnarray}
{\pi \over \alpha_s(s)}
&=&{\pi \over \overline\alpha_s(s)}\,(1+\delta x)
+ { \beta_1 \over \beta_0}\log(1+\delta x)
\nonumber\\
&&+{ \overline\alpha_s(s) \over \pi}
\left(
{ \beta_0\overline\beta_2 - \beta_1^2 \over \beta_0^2}\
{ \delta x \over 1+\delta x}
+{ \beta_1^2 \over \beta_0^2}\
{ \log(1+\delta x) \over 1+\delta x}
- { \delta\beta_2 \over \beta_0}{ 1 \over 1+\delta x}
\right)\nonumber\\
&&+\cdots.
\label{runningalphamod}
\end{eqnarray}
Here,
\begin{equation}
\delta x = \beta_0\, \delta t\,
{ \overline\alpha_s(s) \over \pi}.
\end{equation}

In the framework of this paper (except for the appendix),
light quark masses do
not appear in $R(s)$ because they are set to zero. However, the top
quark mass does appear, starting at order $\alpha_s^2$. Thus it is
necessary to state carefully how we define $m_t$. We let $\overline
m_t(s)$ be the running top quark mass within the $\overline{\rm MS}$
scheme. At the level of perturbation theory at which we work, we need
the one loop evolution of $\overline m_t(s)$, which we write as
\begin{equation}
\overline m_t^2(s) \approx
\overline m_t^2(M_Z^2)\
\exp\!\left(
2\gamma_0 \int^s_{M_Z^2} { d\mu^2 \over \mu^2}\
{ \overline\alpha_s(\mu^2) \over \pi}
\right)
\approx m_t^2(M_Z^2)\
\left(
{ \overline\alpha_s(M_Z^2) \over \overline\alpha_s(s)}
\right)^{2\gamma_0/\beta_0}
\end{equation}
with
\begin{equation}
\gamma_0 = -1.
\end{equation}
(See, for instance, Ref.~\cite{gammaM}). One can, of course, use a
different scheme and define a running mass
\begin{equation}
m_t^2(s) = \overline m_t^2(s)\,[1 + C_1 \overline\alpha_s(s) +\cdots].
\end{equation}
We do so, absorbing the first coefficient $C_1$ into a change of
scale by an amount $\delta t_m$. Thus we define
\begin{equation}
m_t^2(s)
\approx
\overline m_t^2(s)\
\left(
{\overline\alpha_s(s) \over
\overline\alpha_s(s e^{\delta t_m})}
\right)^{2\gamma_0/\beta_0}.
\label{massdef}
\end{equation}
The parameter $\delta t_m$ can be chosen independently from the
scaling parameter $\delta t$ in the definition (\ref{alphasdef}) of
the coupling.

The dependence of ${\cal R}(s)$ on the top quark mass
is quite small, so the dependence of ${\cal R}(s)$ on $\delta t_m$ is
also small. In fact, we find that ${\cal R}(M_Z^2)$ varies by only 0.3
parts per mill for $-4 < \delta t_m < 4$. In order to limit the parameter
space to be explored in our numerical examples, we therefore set
\begin{equation}
\delta t_m = 0.
\end{equation}

Thus the running top mass at $s = M_Z^2$ in our examples is simply the
$\overline{\rm MS}$ running top mass $\overline m_t(M_Z^2)$. We take
$\overline m_t(M_Z^2) = 170.2 {\ \rm GeV}$, which corresponds to a pole
mass of $\tilde m_t = 170 {\ \rm GeV}$ after use of
\begin{equation}
\overline m_t^2(M_Z^2)
\approx
\tilde m_t^2\
\exp\!\left(
-{ 8 \over 3}\, { \overline\alpha_s(\tilde m_t^2) \over \pi}
\right)
\left(
{\overline\alpha_s(\tilde m_t^2) \over \overline\alpha_s(M_Z^2)}
\right)^{2\gamma_0/\beta_0}.
\end{equation}
(See, for instance, Ref.~\cite{gammaM}). The value 170 GeV is estimated
in Ref.~\cite{DESMoriond} from the CDF and D0 results \cite{CDFD0}.

\section{Perturbative expansions}
\label{PerturbativeExpansions}

With the theoretical framework defined in Sec.~\ref{Introduction},
the theoretical expression for $R(s)$ has the form
\begin{equation}
R(s) = R_0\left\{ 1 + {\cal R}(s)\right\}.
\end{equation}
Here $R_0$ is the value of $R(s)$ in the parton model, without
perturbative QCD corrections. The QCD corrections are contained in
${\cal R}(s)$,
\begin{equation}
{\cal R}(s) =
{\cal R}_1\ {\alpha_s(s) \over \pi}
+{\cal R}_2 \left(\alpha_s(s) \over \pi\right)^2
+{\cal R}_3 \left(\alpha_s(s) \over \pi\right)^3
+\cdots.
\label{Rexpansion}
\end{equation}
The value of ${\cal R}(s)$ calculated in finite order perturbation
theory depends on the parameters $\delta t$, $\delta\beta_2$ and $\delta
t_m$ that define the renormalization scheme. We keep these parameters
arbitrary in this analysis in order to be able to test the sensitivity
of the calculated value of ${\cal R}(s)$ to their choice. As already
noted, the dependence of ${\cal R}(s)$ on $\delta t_m$ is negligible.

The $t$ dependence of $\alpha_s(s e^t)$ is given by the
renormalization group equation~(\ref{rengroup}). The coefficients of
the $\beta$ function that appears in this equation are \cite{beta}
\begin{eqnarray}
\beta_0&=&(33 - 2\,N_f)/12,
\nonumber\\
\beta_1&=&(306 - 38\,N_f)/48,
\label{betacoefs}\\
\beta_2&=&(77139 - 15099\,N_f +325\,N_f^2)/3456 + \delta\beta_2\,,
\nonumber
\end{eqnarray}
where $N_f = 5$ is the number of light quark flavors used throughout
this paper.

The coefficients ${\cal R}_1,{\cal R}_2,{\cal R}_3$ are \cite{R2,R3}
\begin{eqnarray}
{\cal R}_1&=&1,
\nonumber\\
{\cal R}_2&=& {365 \over 24} - 11\, \zeta(3)
- N_f \left[{11 \over 12} - {2\,\zeta(3) \over 3}\right]
- { 1 \over \sum_i(v_i^2 + a_i^2)}\
\left[{37 \over 12} + \log\!\left(m_t(s)^2 \over M_Z^2\right)\right]
\nonumber\\
&&+ \beta_0\, \delta t,
\label{Rcoefs}\\
{\cal R}_3&=&
{87029 \over 288} - {1103\, \zeta(3) \over 4}
+ {275\, \zeta(5) \over 6}
- {\beta_0^2 \pi^2 \over 3}
\nonumber\\
&& + N_f \left[
- {7847 \over 216} + {262\, \zeta(3) \over 9}
- {25\, \zeta(5) \over 9}
\right]
 + N_f^2 \left [
{151 \over 162} - {19\, \zeta(3) \over 27}
\right]
\nonumber\\
&&
+ { \left(\sum_i v_i\right)^2 \over \sum_i(v_i^2 + a_i^2)}
\left[
{ 55 \over 72} - { 5\,\zeta(3) \over 3}
\right]
\nonumber\\
&&
+ { 1 \over \sum_i(v_i^2 + a_i^2)}
\left[
-18.65440 - { 31 \over 18}\,\log\!\left(m_t(s)^2\over M_Z^2\right)
+ { 23 \over 12}\,\log^2\!\left(m_t(s)^2\over M_Z^2\right)
+ 2\gamma_0\,\delta t_m
\right]
\nonumber\\
&&
+\beta_0^2\,(\delta t)^2
+\left[\beta_1  + 2\beta_0
{\cal R}_{2,0}\right]\,\delta t
-{ \delta\beta_2 \over  \beta_0}.
\nonumber
\end{eqnarray}
Here $(v_i, a_i)$ with $i = u,d,s,c,b$ is the (vector, axial vector)
coupling of the quark of flavor $i$ to the $Z$ boson, as specified in
the appendix. We use ${\cal R}_{2,0}$ to denote ${\cal R}_{2}$ with
$\delta t = 0$. We recall that  the mass anomalous dimension is
$\gamma_0 = -1$ and that the parameters $\delta t$, $\delta \beta_2$ and
$\delta t_m$ give the scheme dependence, as described in
Sec.~\ref{RunningCoupling}.

The numerical values are
\begin{eqnarray}
\beta_0&\approx& 1.92,
\nonumber\\
\beta_1&\approx& 2.42,
\\
\beta_2&\approx& 2.83 + \delta\beta_2,
\nonumber
\end{eqnarray}
and (with $\delta t_m = 0$ and $m_t(s) = \overline m_t(M_Z^2)$)
\begin{eqnarray}
{\cal R}_1&=&1,
\nonumber\\
{\cal R}_2&\approx&
  0.76
+ 1.92\, \delta t ,
\\
{\cal R}_3&\approx&
- 15.73
+  5.35\, \delta t
+  3.67\, (\delta t)^2
-  0.52\, \delta\beta_2.
\nonumber
\end{eqnarray}

\begin{figure}[htb]
\centerline{ \DESepsf(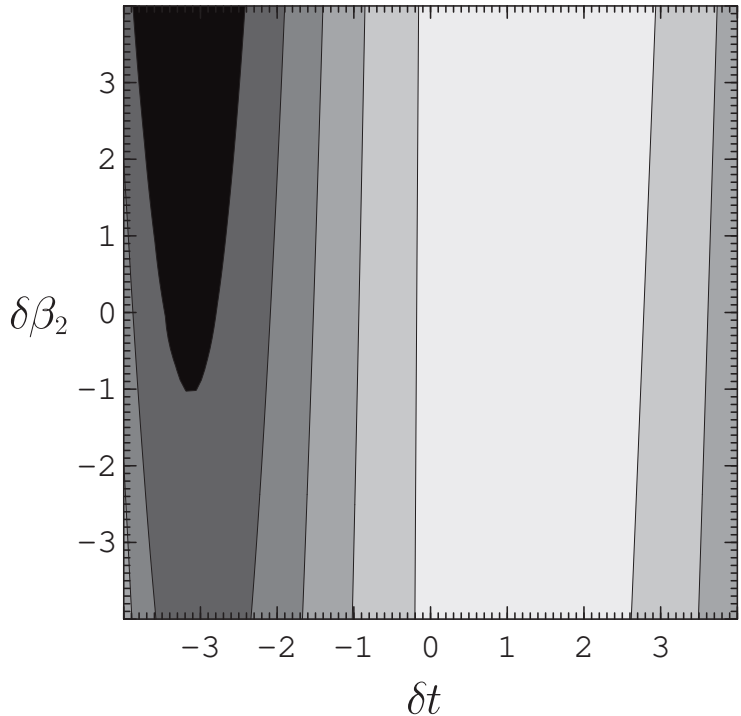 width 10 cm) }
\caption{
Contour plot of the simple third order perturbative approximant ${\cal
R}_A(M_Z^2;{\rm Pert})$, Eq.~(\protect\ref{simpleA}), versus the scheme
fixing parameters $\delta t$ and $\delta\beta_2$, with $\delta t_m =
0$.}
\label{contourRfig}
\end{figure}

In this paper, we will define various approximations ${\cal R}_A$
to ${\cal R}$. The first of these is the simple third order perturbative
approximation:
\begin{equation}
{\cal R}_A(s;{\rm Pert})
= \sum_{j=1}^3 {\cal R}_j \left(\alpha_s(s) \over \pi \right)^j .
\label{simpleA}
\end{equation}
As discussed in Sec.~\ref{RunningCoupling}, the renormalization scheme
ambiguity can provide an estimate, or at least a lower bound, on the
theoretical uncertainty produced by truncating perturbation theory at
order $\alpha_s^3$. To investigate this ambiguity, we show in
Fig.~\ref{contourRfig} a contour plot of ${\cal R}_A(M_Z^2;{\rm Pert})$
as a function of $\delta t$ and $\delta\beta_2$, with $\delta t_m = 0$.
The range shown for the scale parameter, $-4< \delta t<4$, corresponds
to  scales $\mu = [se^t]^{1/2}$ in the range $ 0.14\, M_Z \approx M_Z
e^{-2} < \mu < M_Z e^2 \approx 7.4\, M_Z$ in Eq.~(\ref{alphasdef}). The
range shown for $\delta \beta_2$ corresponds to schemes with $-1.2
\approx \overline \beta_2 - 4 < \beta_2 < \overline\beta_2 + 4 \approx
6.8$.

\begin{figure}[htb]
\centerline{ \DESepsf(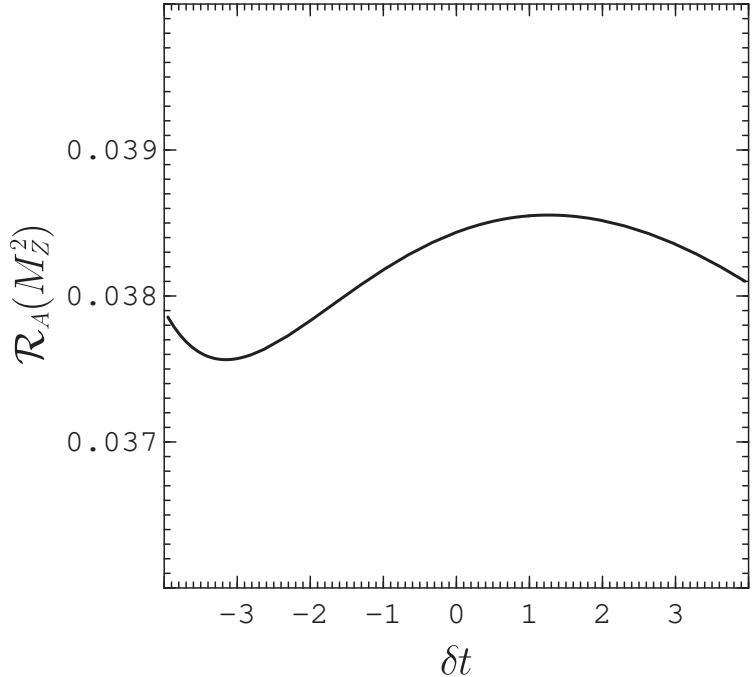 width 10 cm) }
\caption{
Plot of the usual third order perturbative approximant ${\cal
R}_A(M_Z^2;{\rm Pert})$ versus the scheme fixing parameter $\delta
t$ with $\delta\beta_2 = 0$ and $\delta t_m = 0$.}
\label{Rpertvsdt}
\end{figure}

We learn from  Fig.~\ref{contourRfig} that ${\cal R}_A(M_Z^2;{\rm
Pert})$ is not very sensitive to $\delta \beta_2$. Accordingly, we set
$\delta \beta_2 = 0$ and plot ${\cal R}_A(M_Z^2;{\rm Pert})$ versus
$\delta t$ in Fig.~\ref{Rpertvsdt}. We note that ${\cal R}_A(M_Z^2;{\rm
Pert})$ varies by about 2.6\% between its local maximum
and its local minimum. We conclude that ${\cal R}(M_Z^2)$ probably lies
within this 2.6\% range. Thus we ascribe a theoretical error of
$\pm 2.6\%/2 = \pm 1.3\%$ to the value of ${\cal R}(M_Z^2)$. In the
remainder of this paper, we attempt to reduce this error by using more
sophisticated methods than simply taking the first three perturbative
terms in ${\cal R}(M_Z^2)$.

The perturbative series for ${\cal R}(M_Z^2)$ provides our starting
point. We see that the series for $\beta(\alpha_s)$ is nicely
behaved, but that the series for ${\cal R}(s)$ not as well behaved,
with a large value for ${\cal R}_3$ at $\delta t = \delta \beta_2 =
\delta t_m = 0$. In fact, this large value can be attributed to the
term $-\beta_0^2 \pi^2/3 \approx - 12.1$.

\section{$\pi^2$ terms}
\label{PiSquare}

The offending $\pi^2$ term in ${\cal R}_3$ arises, at a rather
mechanical calculational level, because factors of $\log(-s\pm
i\epsilon) = \log(s) \pm i\pi$ occur in the calculation, leading to
powers of $\pi$ in the result.  In order to see what happens at
higher orders of perturbation theory, we write $R(s)$ as a
discontinuity:
\begin{equation}
R(s) = { C \over 2\pi i}
\left\{
\Pi(-s+i\epsilon,\mu^2) - \Pi(-s-i\epsilon,\mu^2)
\right\}.
\label{PitoR}
\end{equation}
Here $C$ is a normalization constant and $\Pi(Q^2,\mu^2)$ is
the standard $Z$ boson self-energy function including the QCD
contribution. It is proportional to the Fourier transform of the time
ordered product of two weak current operators. The current
operators carry momentum $q^\mu$. We define $Q^2 = - q^\mu q_\mu$, so
that $Q^2>0$ if the momentum $q^\mu$ is spacelike. The function $\Pi$
depends on the renormalization scale $\mu^2$. However the function
\begin{equation}
D(Q^2) = -Q^2 { \partial \over \partial Q^2}\Pi(Q^2,\mu^2)
\label{PitoD}
\end{equation}
is a renormalization group invariant. The derivative here
avoids the overall renormalization in $\Pi$. For this reason, it is
standard practice to work with $D(Q^2)$.

We may write the perturbative expansion of $D(Q^2)$ in the form
\begin{equation}
D(Q^2) = D_0\left\{ 1 + {\cal D}(Q^2)\right\},
\end{equation}
where $D_0$ is the value of $D$ in the parton model and where
\begin{equation}
{\cal D}(Q^2) =
{\cal D}_1\ {\alpha_s(Q^2) \over \pi}
+{\cal D}_2 \left(\alpha_s(Q^2) \over \pi\right)^2
+{\cal D}_3 \left(\alpha_s(Q^2) \over \pi\right)^3
+\cdots.
\label{Dexpansion}
\end{equation}
The first three coefficients ${\cal D}_n$ are the same as the
corresponding ${\cal R}_n$ in Eq.~(\ref{Rcoefs}) if one
substitutes $Q^2$ for $s$, except that ${\cal D}_3$ lacks the term
$-\beta_0^2 \pi^2/3$. The numerical values (with $\delta t_m = 0$
and $m_t(s) = \overline m_t(M_Z^2)$) are
\begin{eqnarray}
{\cal D}_1&=&1,
\nonumber\\
{\cal D}_2&\approx&
  0.76
+ 1.92\, \delta t ,
\\
{\cal D}_3&\approx&
- 3.65
+ 5.35\, \delta t
+ 3.67\, (\delta t)^2
- 0.52\, \delta\beta_2.
\nonumber
\end{eqnarray}
If we stay near $\delta t = 0$, this series appears to be quite nicely
behaved. We believe on the basis of general arguments (to be discussed
in the next section) that the coefficients ${\cal D}_n$ will
eventually grow for large $n$. However, that growth is not apparent
in the first three terms.

The function $D(Q^2)$ is calculated using Euclidean quantum field
theory, in which only very weak infrared singularities occur near the
contour of the internal momentum integrations. On the other hand, a
direct calculation of $R(s)$ involves Minkowski momentum integrations
over regions in which various internal particles can go on shell.
Only some delicate cancellations prevent $R(s)$ from being infinite.
Surely $D(Q^2)$ should be better behaved than $R(s)$. This observation
leads to the following
\begin{quote}
{\it Hypothesis 1.} The perturbative expansion of $D(Q^2)$ remains
well behaved beyond the three terms that are known, subject only to
the eventual growth of the ${\cal D}_n$ dictated by the standard
renormalon and instanton ideas.
\end{quote}
We adopt this hypothesis here, although it is criticised in
Ref.~\cite{Altarelli} on the grounds that there could be other sources
of large perturbative coefficients in ${\cal D}(s)$.

We are interested in the observable function $R(s)$.  If we accept
this {\it Hypothesis 1}, then instead of calculating $R(s)$
directly, we should relate it to the nicely behaved function
$D(Q^2)$. From Eqs.~(\ref{PitoR}) and (\ref{PitoD}) we obtain
\begin{equation}
{\cal R}(s) = { 1 \over 2\pi}
\int_{-\pi}^\pi d\theta\ {\cal D}(s e^{i\theta}).
\label{DtoR}
\end{equation}
In the following section, we will deal with the expected large order
behavior of the ${\cal D}_n$ by following the standard practice of
using the Borel transform $\tilde{\cal D}$ of ${\cal D}$:
\begin{equation}
{\cal D}(Q^2) =
\int_0^{\infty}\! dz\ \exp\!\left(-{\pi z
\over\alpha_s(Q^2)}\right)\
\tilde{\cal D}(z).
\label{borel}
\end{equation}
If we write the perturbative expansion of $\tilde{\cal
D}(z)$ as
\begin{equation}
\tilde{\cal D}(z) =
\tilde{\cal D}_0
+\tilde{\cal D}_1\ z
+\tilde{\cal D}_2\ z^2+\cdots,
\label{tildeDexpansion}
\end{equation}
then
\begin{equation}
\tilde{\cal D}_n = { {\cal D}_{n+1} \over n!}.
\end{equation}
Because of the $1/n!$ factor, the perturbative expansion of
$\tilde{\cal D}$ in powers of $z$ is much nicer than that of
$\tilde{\cal D}$ in powers of $\alpha_s/\pi$. In fact, one expects
$\tilde{\cal D}(z)$ to be analytic near $z=0$. As discussed, for
example, in Ref.~\cite{Mueller1}, there are singularities expected in
the complex $z$ plane, including some on the integration contour along
the positive $z$-axis. In addition $\tilde{\cal D}(z)$ is not expected
to be well behaved as $z \to \infty$. Thus the meaning of the
integration in Eq.~(\ref{borel}) is ambiguous. In this section, we
simply leave it as ambiguous.

We can relate ${\cal R}(s)$ to $\tilde{\cal D}$ by inserting
Eq.~(\ref{borel}) into Eq.~(\ref{DtoR}):
\begin{equation}
{\cal R}(s) =
\int_0^{\infty}\! dz\
\exp\!\left(-{\pi z \over\alpha_s(s)}\right)\
F(\alpha_s(s),z)\
\tilde{\cal D}(z),
\label{tildeDtoR}
\end{equation}
where
\begin{equation}
F(\alpha_s,z) = { 1 \over 2\pi}
\int_{-\pi}^\pi d\theta\
\exp\!\left(- zG(\alpha_s,\theta)\right),
\label{Fdef}
\end{equation}
with
\begin{equation}
G(\alpha_s(s),\theta) =
{\pi \over\alpha_s(s e^{i\theta})}
-{\pi \over\alpha_s(s)}.
\end{equation}
Eq.~(\ref{tildeDtoR}) is the basis for the analysis in this paper. We
note that the factor $\pi/\alpha_s(s)$ in the exponent in
Eq.~(\ref{tildeDtoR}) is big, about 30 for $\alpha_s
\approx 0.12$. Therefore the integral over $z$ is dominated by small
$z$, $\beta_0 z\lesssim \beta_0 \alpha_s/\pi \approx 0.06$. Thus we
will be primarily concerned with the expansion of $\tilde{\cal D}(z)$
in powers of $z$.

Before addressing $\tilde{\cal D}(z)$, however, we need a good
approximation for $F(\alpha_s,z)$. Since small $z$ is important, we
are particularly interested in the small $z$ region. However, it is
rather easy to find an approximation for $F(\alpha_s,z)$ that is good
for a wide range of $z$, based on the smallness of its argument
$\alpha_s$. We use the solution (\ref{runningalpha})
of the renormalization group equation~(\ref{rengroup}) for
$\pi/\alpha_s$ in order to derive an approximation for
$G(\alpha_s,\theta)$. We find $G(\alpha_s,\theta) \approx
G_A(\alpha_s,\theta)$ where
\begin{eqnarray}
G_A(\alpha_s,\theta)&=&i\beta_0 \theta
+ { \beta_1 \over \beta_0}\log(1+y)
\nonumber\\
&&+{ \alpha_s \over \pi}
\left(
{ \beta_0\beta_2 - \beta_1^2 \over \beta_0^2}\
{ y \over 1+y}
+{ \beta_1^2 \over \beta_0^2}\
{ \log(1+y) \over 1+y}
\right),
\end{eqnarray}
where
\begin{equation}
y = i\beta_0\theta\,{ \alpha_s \over \pi}.
\end{equation}
Further terms in this series involve higher powers of $\alpha_s/\pi$
times functions of $y$ that vanish for $y \to 0$. The ordinary
perturbative expansion of $G(s)$ results from expanding in powers of $y$,
which is proportional to $\alpha_s$, then omitting terms beyond $y^2$ or
$y\alpha_s$. However, $\alpha_s(s)/\pi \approx 1/30$ while $|y|\approx
1/5$ at $\theta = \pi$. Thus $\alpha_s(s)/\pi$ is a much better
expansion parameter than $y$. Since we don't have to expand in $y$, we
don't.

\begin{figure}[htb]
\centerline{ \DESepsf(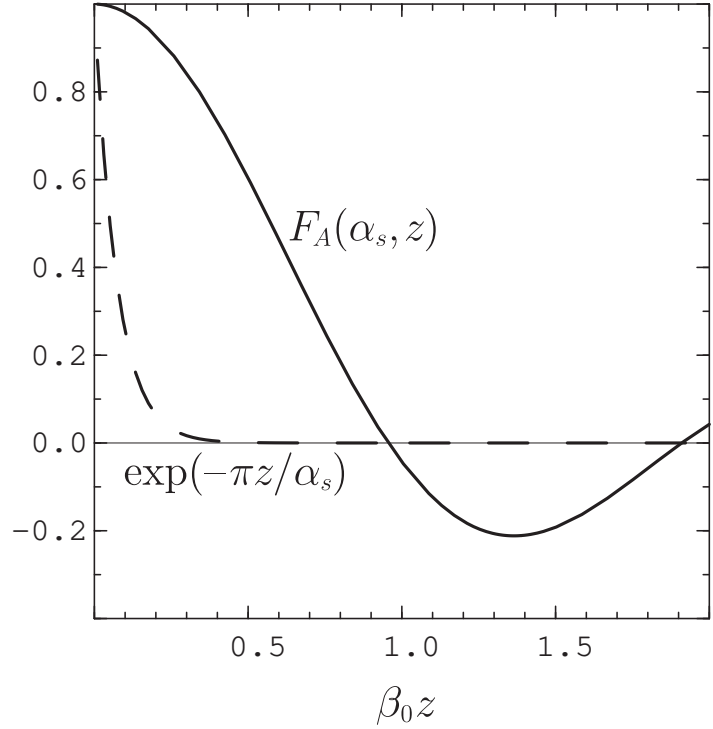 width 10 cm) }
\caption{
Graph of $F_A(\alpha_s,z)$ and $\exp(- \pi z/\alpha_s)$ versus
$\beta_0 z$ with $\alpha_s = 0.12$
}
\label{FAfig}
\end{figure}

We now have an approximation  $G_A(\alpha_s,\theta)$ for
$G(\alpha_s,\theta)$. Our corresponding approximation
$F_A(\alpha_s,z)$ for $F(\alpha_s,z)$ is
\begin{equation}
F_A(\alpha_s,z) = { 1 \over 2\pi}
\int_{-\pi}^\pi d\theta\
\exp\!\left(- z G_A(\alpha_s,\theta)\right),
\label{FA}
\end{equation}
with the integral computed to sufficient accuracy by numerical
methods. In Fig.~\ref{FAfig} we show a graph of $F_A(\alpha_s,z)$
versus $\beta_0 z$ superimposed on a graph of $\exp(- \pi
z/\alpha_s)$, all with $\alpha_s = 0.12$.

How good is our approximation $F_A(\alpha_s,z)$? The first omitted
term in $G(\alpha_s,\theta)$ is, in the $\overline {\rm MS}$ scheme,
\begin{equation}
\Delta G(\alpha_s) = \left( \alpha_s \over \pi \right)^2
h(x),
\end{equation}
where
\begin{eqnarray}
h(x) &=&
-{\beta_1^3 \over 2\beta_0^3}
{ \log^2(1+x) \over (1+x)^2}
+{\beta_1\beta_2 \over \beta_0^2}
{ \log(1+x) \over (1+x)^2}
\nonumber\\
&&  +
{ \beta_1^3 \over 2\beta_0^3}
{ x^2 \over  (1+x)^2}
-
{\beta_1\beta_2 \over \beta_0^2}
{ x \over (1+x)}
 +
{\beta_3\over 2\beta_0}
{ x(x+2) \over (1+x)^2}.
%
\end{eqnarray}
This term contains a factor $(\alpha_s/\pi)^2 \approx 10^{-3}$ for
$\alpha_s \approx 0.1$. This factor multiplies $h(x)$, which cannot
be evaluated because it contains the unknown coefficient $\beta_3$.
However, we can see from the structure of $h(x)$ that it is not large
unless $\beta_3$ is large. We can get a quantitative idea of the
effect of $\Delta G$ by choosing some plausible values for $\beta_3$
and then calculating ${\cal R}(s)$ with $\Delta G$ included. We find
that, taking $\beta_3$ in the range $-10 < \beta_3 < 10$,  the
fractional change ${\cal R}(M_Z^2)$ induced by including $\Delta
G(\alpha_s)$ is no larger than $2\times 10^{-5}$. Since this error is
small compared to our target error of a few per mill, we can safely
neglect it.

We thus obtain an approximation for $\cal R$ that uses third order
perturbation theory but sums certain ``$\pi^2$'' effects to all
orders:
\begin{equation}
{\cal R}_A(s;\pi^2) =
\int_0^{\infty}\! dz\
\exp\!\left(-{\pi z \over\alpha_s(s)}\right)\
F_A(\alpha_s(s),z)\
\tilde{\cal D}_A(z),
\label{pisqA}
\end{equation}
where $F_A$ is given in Eq.~(\ref{FA}) and $\tilde{\cal D}_A(z)$ is
simply $\tilde{\cal D}(z)$, Eq.~(\ref{tildeDexpansion}), expanded to
second order in $z$.

This treatment of $\pi^2$ terms is similar in spirit to that of Le
Diberder and Pich \cite{DiberderPich}, who expand ${\cal D} (s
e^{i\theta})$ in Eq.~(\ref{DtoR}) in perturbation theory, use
Eq.~(\ref{runningalpha}) for $\alpha_s(s e^{i\theta})$, and perform the
$\theta$ integral exactly. We simply embed this approach into the Borel
transform.

\begin{figure}[htb]
\centerline{ \DESepsf(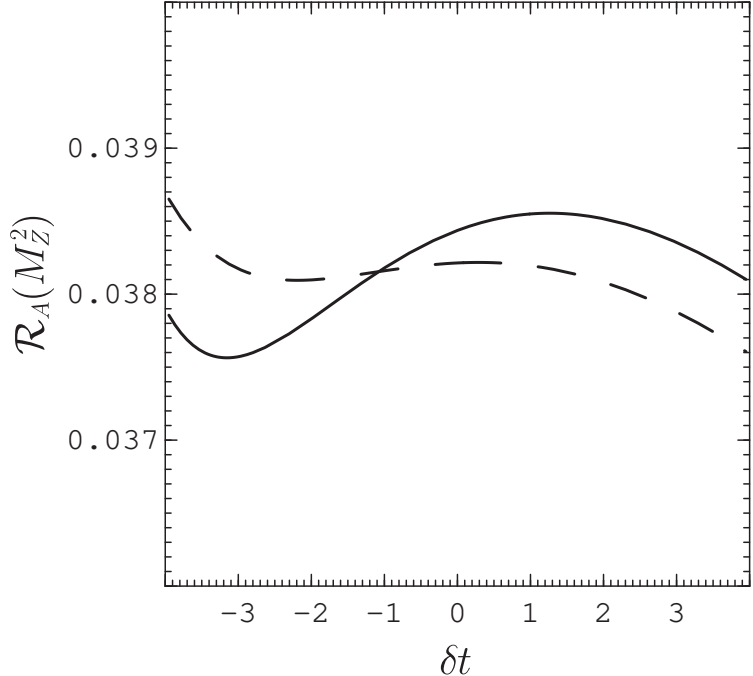 width 10 cm) }
\caption{
Plot of $\pi^2$-summed approximant ${\cal R}_A(M_Z^2;\pi^2)$,
Eq.~(\protect\ref{pisqA}), (dashed line), versus the scheme fixing
parameter $\delta t$ with $\delta\beta_2 = 0$ and $\delta t_m = 0$. We
also show ${\cal R}_A(M_Z^2;{\rm Pert})$ from
Fig.~\protect\ref{Rpertvsdt} (full line).}
\label{Rpisqvsdt}
\end{figure}

In Fig.~\ref{Rpisqvsdt} we plot ${\cal R}_A(M_Z^2;\pi^2)$ versus the
scheme parameter $\delta t$ with the other scheme parameters set to
$\delta\beta_2 = 0$ and $\delta t_m = 0$. We overlay the plot of
${\cal R}_A(M_Z^2;{\rm Pert})$ from  Fig.~\ref{Rpertvsdt}. We note that
${\cal R}_A(M_Z^2;\pi^2)$ varies by about 0.32\% between its local
maximum and its local minimum. This is a much smaller variation than
that of ${\cal R}_A(M_Z^2;{\rm Pert})$. A very optimistic view would be
that ${\cal R}(M_Z^2)$ probably lies within this 0.32\% range, so that
one would ascribe a theoretical error of $\pm 0.32\%/2 = \pm 0.16\%$ to
the value of ${\cal R}(M_Z^2)$. However, this error estimate is smaller
than other error estimates that we will develop later. We therefore
regard the flatness of the curve for ${\cal R}_A(M_Z^2;\pi^2)$ as being
partially the result of an accidental cancellation, and refrain from
taking 0.16\% as a reasonable error estimate.

We close this section by emphasizing the observation that
the straightforward perturbative expansion of ${\cal R}(s)$ is, in
part, an expansion in powers of $\beta_0\theta\,[{ \alpha_s(s) /
\pi}]$, with $\theta \sim \pi$, instead of an expansion in powers of
$[\alpha_s(s)/\pi] \approx 1/30$. One can attribute the appearance
of ``$\pi^2$'' terms in ${\cal R}(s)$ to this phenomenon. This
observation helps to make {\it Hypothesis 1} plausible.
Unfortunately, this argument is only suggestive, since one can not
be sure that there are not ``bad expansion parameters'' lurking
somewhere in the calculation of ${\cal D}(Q^2)$. In the next
section, we turn to the behavior of the perturbative coefficients in
$\tilde{\cal D}(z)$, assuming that the evidence for a bad
expansion parameter is not, in fact, lurking just beyond the last
calculated coefficient.

\section{Truncation of the integral}
\label{Truncation}

If we do not expand $\tilde{\cal D}(z)$ in powers of $z$, then at
this point we have an approximation for
$\cal R$ of the form
\begin{equation}
{\cal R}(s) \approx
\int_0^{\infty}\! dz\
\exp\!\left(-{\pi z \over\alpha_s(s)}\right)\
F_A(\alpha_s(s),z)\
\tilde{\cal D}(z),
\end{equation}
with $F_A$ given in Eq.~(\ref{FA}). Since $\alpha_s(s)$ is small, the
dominant integration region is $z \ll 1$. Indeed, taking $\alpha_s
\approx 0.12$, we have $\exp\!\left(-{\pi z /\alpha_s}\right) <
10^{-3}$ for $\beta_0 z > 0.51$. Thus it is useful to write ${\cal
R}$ in the form
\begin{equation}
{\cal R}(s) \approx
\int_0^{z_{\rm max}}\! dz\
\exp\!\left(-{\pi z \over\alpha_s(s)}\right)\
F_A(\alpha_s(s),z)\
\tilde{\cal D}(z)
+ {\cal R}_R(s)
\label{inversion2}
\end{equation}
with $\beta_0 z_{\rm max} \gtrsim 0.5$. The fundamental question
of how the ``sum of perturbation theory'' is precisely defined relates
to the definition of ${\cal R}_R$. In turn, this question is related
to how the renormalon and instanton singularities are treated and to
the question of the convergence of the integral at large $z$. However,
our purpose here is at once more modest and more practical. We adopt
\begin{quote}
{\it Hypothesis 2.} It is safe to ignore the large $z$ part of the
Borel integral when calculating ${\cal R}(s)$ for $s \sim M_Z$,
even though this part of the integral is ill defined.
\end{quote}
We thus neglect ${\cal R}_R$ and concentrate on the integral up to
$z_{\rm max}$ in Eq.~(\ref{inversion2}). The advantage is that we
can use approximations for $\tilde{\cal D}(z)$ that have
singularities on the positive $z$ axis outside of the region of
integration.

We can test the sensitivity of the computed value of $\cal R$
to $z_{\rm max}$ by replacing $\tilde{\cal D}(z)$ by its second order
expansion in powers of $z$. Then the ratio of the
two terms in Eq.~(\ref{inversion2}) with $z_{\rm max} = 0.5$ (for $s =
M_Z^2,\ \delta t = \delta t_m = \delta \beta_3 = 0$) is ${\cal
R}_R(s)/ {\cal R}_A(s;\pi^2) \approx 6\times 10^{-4}$.

\pagebreak
\section{Accounting for renormalons}
\label{Renormalons}

We now turn to the perturbative expansion
\begin{equation}
\tilde{\cal D}(z) = \sum_0^\infty \tilde{\cal D}_n\, z^n.
\end{equation}
The coefficients $\tilde{\cal D}_n$ can be expressed as an
integral,
\begin{equation}
\tilde{\cal D}_n = { 1 \over 2 \pi i}
\int_{\cal C} d z\ z^{-n-1}\tilde{\cal D}(z).
\end{equation}
The contour $\cal C$ encloses the point $z = 0$ but excludes any
singularities of $\tilde{\cal D}(z)$. Thus the behavior of the
${\cal D}_n$ at large order $n$ is controlled by the part of the
contour that lies nearest to $z=0$, which in turn is controlled by
the singularities of $\tilde {\cal D}(z)$ that are nearest to
$z=0$. A singularity of the form $(z - z_0)^{-A}$ makes a
contribution to ${\cal D}_n$ that is proportional to $z_0^{-n}\,
n^{A-1}$. Thus the most important determinant of the singularity's
contribution to the $\tilde{\cal D}_n$ at large $n$ is its location,
$z_0$. A small $z_0$ produces large coefficients. The next most important
determinant is the strength of the singularity, $A$. A large positive
value of $A$ produces large coefficients.

The nearest singularities are thought to be the first two ultraviolet
renormalon singularities at $\beta_0 z = - 1$ and $- 2$ and the first
infrared renormalon singularity at $\beta_0 z=+2$ \cite{Mueller1,Mueller2}.
In this section, we use the available information on these
singularities to obtain a perturbative expansion that has better
convergence properties. Of course, ``better convergence properties''
refers to the perturbative coefficients for large $n$. It is
problematical whether convergence improvement helps already after only
three terms of the series.

The first ultraviolet renormalon singularity is at $\beta_0 z=-1$. This
is the singularity that is closest to the origin (at least so far as
anyone knows). It thus controls the large order behavior of the
perturbative series. Unfortunately, the theory of the ultraviolet
renormalon singularities is not as simple or as well developed as that for
the infrared renormalon singularities. (See, however,
Ref.~\cite{Zakharov}). For instance, the strength of the singularity is
not known.

The first infrared renormalon singularity is at $\beta_0 z=+2$. There
are other singularities farther away from the origin along the positive
real $z$-axis, but we need not be concerned with them: since they lie
farther from $z = 0$, their contribution to the large order behavior of
the perturbative coefficients is weaker than that of the first
singularity. It is significant that there is no infrared renormalon
singularity at $\beta_0 z = +1$. The first infrared renormalon
singularity has a power behavior,
\begin{equation}
\tilde{\cal D}(z) \sim
 c  \left[1-{\beta_0 z\over 2}\right]^{-1 - 2\beta_1/\beta_0^2},
\end{equation}
where $c$ is a constant \cite{Mueller1,Mueller2}. Numerically, the
exponent is ${-1 - 2\beta_1/\beta_0^2} \approx - 2.3$.

We can make use of this information. Consider the function
\begin{equation}
\tilde{\cal C}(z) =
\tilde{\cal D}(z)
\left[1-{\beta_0 z\over 2}\right]^{1 + 2\beta_1/\beta_0^2}.
\label{tildecalCdef}
\end{equation}
The factor multiplying $\tilde{\cal D}(z)$ cancels its divergence
as $\beta_0 z \to 2$. The function $\tilde{\cal C}(z)$ is still singular
at $\beta_0 z = 2$, since if we multiply a term in $\tilde{\cal D}(z)$
that is analytic at $\beta_0 z = 2$ by the nonanalytic factor, we
create a nonanalytic term. However, singularity is much weaker than
it was, behaving like
\begin{equation}
\tilde{\cal C}(z) \sim
c \left[1-{\beta_0 z\over 2}\right]^{1 + 2\beta_1/\beta_0^2}.
\end{equation}
Thus the perturbative expansion of $\tilde{\cal C}(z)$ would be
better behaved than that of $\tilde{\cal D}(z)$ at large order if
it were not for the fact that the leading ultraviolet renormalon
singularity at $\beta_0 z = -1$ dominates the large order behavior.

We can, however, improve the large order behavior arising from the
leading ultraviolet renormalon by merely moving it out of the way by
means of a good choice of variable. Following Mueller
\cite{Mueller2} we define a new variable $\zeta$ by
\begin{equation}
\beta_0 z = { \beta_0\zeta \over (1 - \beta_0\zeta/4)^2}
\ ,\hskip 1 cm
\beta_0 \zeta = 4\,{ \sqrt{1 + \beta_0z} - 1 \over
          \sqrt{1 + \beta_0z} + 1 }.
\end{equation}
This transformation maps the origin of the $z$-plane onto the origin of
the $\zeta$-plane. We have chosen the normalization of $\zeta$ such that
\begin{equation}
\zeta \sim z + {\cal O}(z^2).
\end{equation}
near $z = 0$. The map treats specially the interval $\beta_0 z<-1$ on
the negative $z$-axis that contains the ultraviolet renormalon
singularities. The whole complex $z$-plane except for this interval
is mapped to the interior of the disk $|\beta_0\zeta| < 4$ in the
$\zeta$-plane. The singularity-free interval $-1 <\beta_0 z<0$ in the
negative $z$-axis is mapped onto the interval $-4<\beta_0\zeta<0$ of
the negative $\zeta$-axis while the interval $0<\beta_0 z<\infty$ on
the positive $z$-axis, which contains the infrared renormalon and
instanton singularities, is mapped into the interval $0 <
\beta_0\zeta < 4$ of the positive $\zeta$-axis.

We consider the function
\begin{equation}
\tilde{\cal B}(\zeta) = \tilde{\cal C}(z(\zeta)).
\label{tildecalBdef}
\end{equation}
The singularity of $\tilde{\cal B}(\zeta)$ that is nearest to the
origin of the $\zeta$-plane is the first infrared renormalon
singularity, which is at
\begin{equation}
\beta_0\zeta =  4{ \sqrt 3 - 1 \over \sqrt 3
+ 1} \approx 1.1.
\end{equation}
Thus moving the ultraviolet renormalon singularity away has had a
price. We have moved the infrared renormalon singularity closer to
the origin. However, we have previously softened the infrared
renormalon singularity, so the price is not too great. The net
effect should be an improvement.

The effect of singularity mapping has been investigated recently by
Altarelli {\it et al.} \cite{Altarelli}. However, these authors did
not also soften the infrared renormalon singularity. They found that
there was no gain in this method.

In order to use the singularity softening and mapping, we use the
first three terms in the perturbative expansion of $\tilde{\cal D}(z)$,
\begin{eqnarray}
\tilde{\cal D}(z) &=& 1
+ \biggl[\,
  0.40
+ 1.00\, \delta t
\biggr]\,(\beta_0 z)
\nonumber\\
&&\ \
+\ \biggl[
- 0.50
+ 0.73\, \delta t
+ 0.50\, (\delta t)^2
\biggr]\,(\beta_0 z)^2 + \cdots,
\label{tildecalDterms}
\end{eqnarray}
to calculate the first three terms in the
perturbative expansion of $\tilde{\cal B}(\zeta)$. The result is
\begin{eqnarray}
\tilde{\cal B}(\zeta) &=& 1
+ \biggl[
- 0.76
+ 1.00\, \delta t
\biggr]\,(\beta_0 \zeta)
\nonumber\\
&&\ \
+\ \biggl[
- 0.96
+ 0.07\, \delta t
+ 0.50\, (\delta t)^2
\biggr]\,(\beta_0 \zeta)^2 + \cdots.
\label{tildecalBterms}
\end{eqnarray}
(Here we have displayed the coefficients numerically, with the
choices $\delta \beta_2 = \delta t_m = 0$ and $m_t(s) = \overline
m_t(M_Z^2)$.)

This perturbative series for $\tilde{\cal B}(\zeta)$ is
supposed to be better behaved at large orders than was the
perturbative series for $\tilde{\cal D}(z)$. The expected improvement
is not, however, visible in the first three terms. In fact, we
started with a series that was quite well behaved, and we have
applied a rather mild improvement program. As long as the infrared
and untraviolet renormalon singularities are as described in this
section, this program may be expected to make the perturbative
coefficients smaller at high order, but one cannot expect too much to
happen at order two.

An example of this procedure applied to a simple model may be useful
as an illustration of what happens at high order. Suppose that
\begin{equation}
\tilde {\cal D}(z) =
{ z \over \left[ 1+\beta_0 z\right]}
+{ 1 \over \left[ 1-\beta_0 z/2\right]^p}
\end{equation}
with $p = {1 + 2 \beta_1/ \beta_0^2}$. Then the perturbative
expansion of $\tilde {\cal D}(z)$ is
\begin{eqnarray}
\tilde {\cal D}(z)&=&
1
+  1.68\,  \beta_0 z
+  0.44\, (\beta_0 z)^2
+  1.21\, (\beta_0 z)^3
\nonumber\\&&
-  0.06\, (\beta_0 z)^4
+  0.81\, (\beta_0 z)^5
-  0.35\, (\beta_0 z)^6 + \cdots.
\end{eqnarray}
Applying the renormalon improvement procedure gives the function
$\tilde {\cal B}(\zeta)$ with a perturbative expansion
\begin{eqnarray}
\tilde {\cal B}(\zeta) &=&
1
+  0.52\, \beta_0\zeta
-  0.87\, (\beta_0\zeta)^2
+  0.30\, (\beta_0\zeta)^3
\nonumber\\&&
-  0.02\, (\beta_0\zeta)^4
+  0.06\, (\beta_0\zeta)^5
-  0.01\, (\beta_0\zeta)^6 + \cdots.
\end{eqnarray}
The series for $\tilde {\cal B}$ is clearly better behaved at high
orders than the series for $\tilde {\cal D}$ . One might claim to see
an improvement beginning with the fourth term, which corresponds to
the first uncalculated term in the case of the real $\tilde {\cal D}$
and $\tilde {\cal B}$ functions. However, at this quite low order of
expansion, the improvement is marginal.

\begin{figure}[htb]
\centerline{ \DESepsf(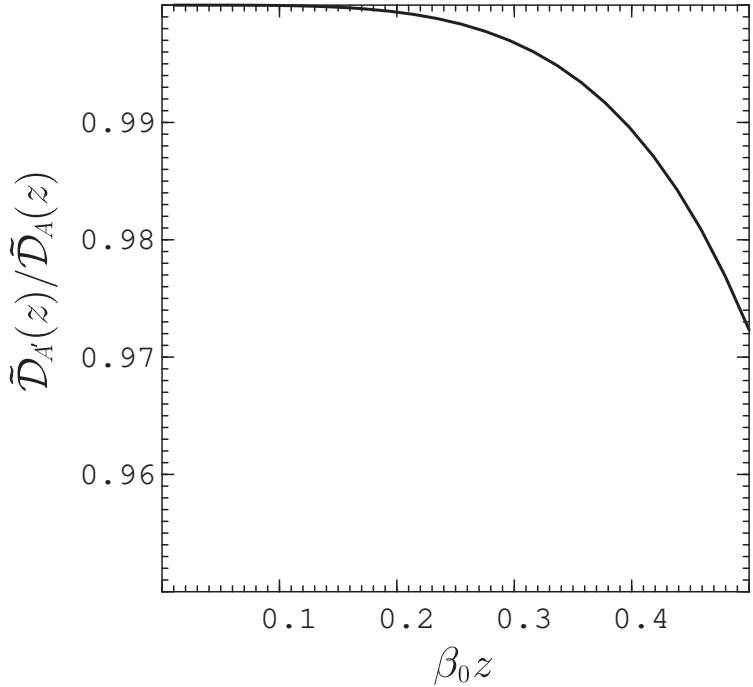 width 10 cm) }
\caption{
Modification of the Borel integrand to account for renormalons.
We plot $\tilde{\cal D}_{A'}(z)/ \tilde{\cal D}_{A}(z)$,
Eqs.~(\protect\ref{Dfctn}) and (\protect\ref{Dmodfctn}), versus
$\beta_0 z$. We set $s = M_Z^2$ and choose the scheme fixing parameters
$\delta t=\delta\beta_2 =\delta t_m = 0$.
}
\label{DoverDmod}
\end{figure}

The procedure for singularity softening and mapping may be
summarized as follows. We calculate the first $N$ terms in the
expansion of $\tilde {\cal B}(\zeta)$ according to
Eqs.~(\ref{tildecalCdef}) and (\ref{tildecalBdef}), where for us $N=
3$. Then we instead of using
\begin{equation}
\tilde{\cal D}_A(z) \equiv \sum_{n=0}^2 \tilde{\cal D}_n z^n
\label{Dfctn}
\end{equation}
for $\tilde{\cal D}(z)$ in Eq.~(\ref{inversion2}), we use
\begin{equation}
\tilde{\cal D}_{A'}(z) \equiv
\left[1-{\beta_0 z \over 2}\right]^{-1 - 2\beta_1/\beta_0^2}\,
\sum_{n=0}^2 \tilde{\cal B}_n\ \left[\zeta(z)\right]^n.
\label{Dmodfctn}
\end{equation}
This gives an approximation for ${\cal R}(s)$ that we may call
${\cal R}_A(s;\pi^2,{\rm R\hbox{'}lons})$:
\begin{equation}
{\cal R}_A(s;\pi^2,{\rm R\hbox{'}lons}) =
\int_0^{z_{\rm max}}\! dz\
\exp\!\left(-{\pi z \over\alpha_s(s)}\right)\
F_A(\alpha_s(s),z)\
{ \sum_{n=0}^2 \tilde{\cal B}_n\ \left[\zeta(z)\right]^n
\over
\left[1-{\beta_0 z / 2}\right]^{1 + 2\beta_1/\beta_0^2}\,}.
\label{Rlons}
\end{equation}
The replacement of $\tilde{\cal D}_{A}(z) $ by $\tilde{\cal
D}_{A'}(z)$ does not modify the integrand much. In
Fig.~\ref{DoverDmod}, we show the ratio $\tilde{\cal D}_{A'}(z)/
\tilde{\cal D}_{A}(z)$ as a function of $\beta_0 z$. We see that
this ratio is nearly 1.0 in the important integration region
$\beta_0 z < 0.2$.

\section{Results}
\label{Results}

\begin{figure}[htb]
\centerline{ \DESepsf(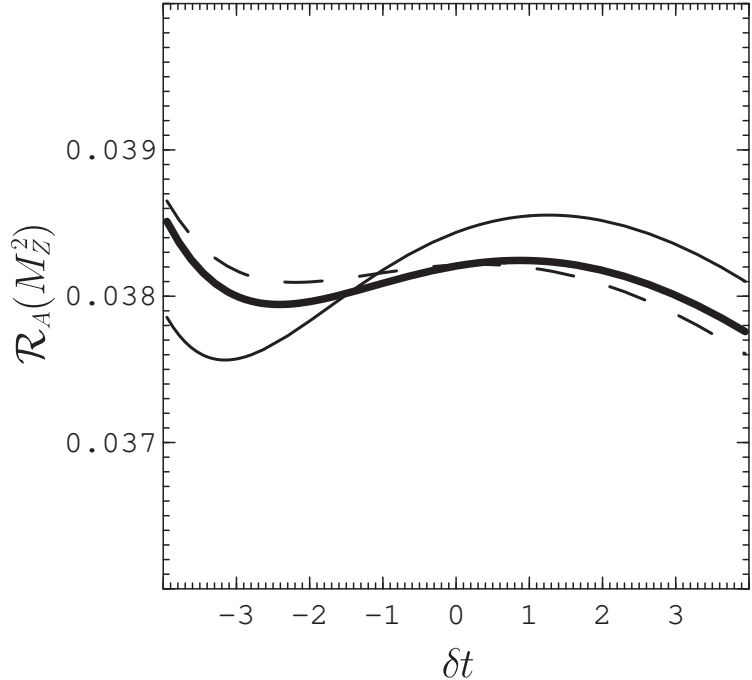 width 10 cm) }
\caption{
Plot of approximant ${\cal R}_A(s;\pi^2,{\rm R\hbox{'}lons})$,
Eq.~(\protect\ref{Rlons}), versus the scheme fixing parameter
$\delta t$ with $\delta\beta_2 = 0$ and $\delta t_m = 0$ (heavy line). We
also show ${\cal R}_A(M_Z^2;{\rm Pert})$ from
Fig.~\protect\ref{Rpertvsdt} (light line) and ${\cal R}_A (M_Z^2;\pi^2)$
from Fig.~\protect\ref{Rpisqvsdt} (dashed line).}
\label{Rnetvsdt}
\end{figure}

We have developed an approximation to ${\cal R}(s)$ that takes
$\pi^2$ contributions into account and uses information about the
leading renormalon singularities to try to improve the convergence of
the perturbative expansion for $\tilde{\cal D}(z)$. In
Fig.~\ref{Rnetvsdt} we plot this approximation, ${\cal
R}_A(s;\pi^2,{\rm R\hbox{'}lons})$, versus the scheme parameter
$\delta t$ with the other scheme parameters set to $\delta\beta_2
= \delta t_m = 0$. We overlay the plots of the pure perturbative
function, ${\cal R}_A(M_Z^2;{\rm Pert})$, and the approximation that
simply takes $\pi^2$ contributions into account, ${\cal R}_A
(M_Z^2;\pi^2)$. We note that ${\cal R}_A(s;\pi^2,{\rm R\hbox{'}lons})$
varies by about 0.8\% between its local maximum and its local minimum.
This suggests that ${\cal R}(M_Z^2)$ probably lies within this 0.8\%
range, so that one would ascribe a theoretical error of $\pm 0.8\%/2 =
\pm 0.4\%$ to the value of ${\cal R}(M_Z^2)$.

We can take another approach to error estimation. We note that the
first three coefficients of $(\beta_0 \zeta)^n$ in
Eq.~(\ref{tildecalBterms}) are all of order 1. That the coefficients
do not appear to be growing or shrinking with $n$ is normal
since the series is expected to have a radius of convergence of about
1 in the variable $\beta_0 \zeta$. We thus expect that the
uncalculated coefficient of $(\beta_0 \zeta)^3$ will also be of
order 1. If we add a term $1\times(\beta_0 \zeta)^3$ to the series
in Eq.~(\ref{Rlons}), ${\cal R}_A(M_Z^2)$ changes by an amount
$\delta {\cal R}_A$ that can serve as an error estimate. We find
$
{ \delta{\cal R}_A / {\cal R}_A(M_Z^2;\pi^2,{\rm R\hbox{'}lons})}
\approx 0.2\%.
$

We thus have three error estimates. From the $\delta t$ dependence
of ${\cal R}_A(M_Z^2;\pi^2)$ we estimated a 0.16\% error. From
consideration of the likely size of the next term in $\tilde {\cal
B}(\zeta)$ we estimated a 0.2\% error. From the $\delta t$ dependence
of ${\cal R}_A(M_Z^2;\pi^2,{\rm R\hbox{'}lons})$ we estimated a 0.4\%
error. We take the largest of these values, 0.4\%, as a reasonable
estimate of the theoretical error (in the spirit of a ``1 $\sigma$''
error).

For the central value, we take the value of ${\cal R}_A (s;\pi^2,{\rm
R\hbox{'}lons})$ at $\delta t = 0$, which is almost exactly also the
value of ${\cal R}_A(s;\pi^2)$ at $\delta t = 0$. This value is
\begin{equation}
{\cal R}_A(M_Z^2;\pi^2,{\rm R\hbox{'}lons})_{\delta t = 0}
\approx (1 - 0.006) \times
{\cal R}_A(M_Z^2;{\rm Pert})_{\delta t = 0}.
\end{equation}
That is, our best estimate for $\cal R$ is renormalized down by
0.6\% compared to the standard $\overline {\rm MS}$ value with a
scale choice $\mu = M_Z$.

One often uses a measurement of ${\cal R}(M_Z^2)$ to extract a value
of $\overline\alpha_s(M_Z^2)$. Recall that, to a good approximation,
${\cal R}(M_Z^2) \propto \overline\alpha_s(M_Z^2)$. Thus the value of
$\overline\alpha_s(M_Z^2)$ extracted from data using the ``standard''
$\overline {\rm MS}$ expression for $\cal R$ (with a scale choice $\mu
= M_Z$) would be renormalized {\it up} by 0.6\% if one uses the
``improved'' version of $\cal R$ presented here:
\begin{equation}
[\overline\alpha_s(M_Z^2)]_{\rm improved} \approx 1.006\
[\overline\alpha_s(M_Z^2)]_{\rm standard}.
\label{shift}
\end{equation}
The fractional error to be ascribed to $\overline\alpha_s(M_Z^2)$ from
uncertainties in the QCD perturbation theory is just the fractional
error in ${\cal R}_A(M_Z^2;\pi^2,{\rm R\hbox{'}lons})$ estimated
above as 0.4\%. This is one third of the 1.3\% error that we would
ascribe to $\overline\alpha_s(M_Z^2)$ extracted using the standard
perturbative approximant ${\cal R}_A(M_Z^2;{\rm Pert})$. The shift
in Eq.~(\ref{shift}) is about the same size as the estimated
theoretical error, so it is marginally significant.

We note that the experimental error for the extraction of $\alpha_s$ by
this method is about 5\% \cite{Rexperiment}, much larger than the QCD
theoretical error that we estimate above. There are also sources of
theoretical error not associated with QCD. According to the estimates of
Hebbeker, Martinez, Passarino and Quast \cite{hebbeker}, the most
important of these are a $\pm 2\%$ uncertainly from electroweak
corrections and a $\pm 2\%$ uncertainly from not knowing the Higgs boson
mass.

\acknowledgments

It is a pleasure to thank  V.~Braun, L.~Clavelli, P.~Coulter,
Z.~Kunszt, A.~Mueller and P.~Raczka for helpful conversations. This work
was supported by the U.S. Department of Energy under grants
DE-FG06-85ER-40224 and DE-FG05-84ER-40141.

\appendix

\section*{Present status of perturbative QCD evaluation
                   of Z decay rates}

The decay rate of the $Z$ boson into quark antiquark pair can be
written in the following form:
\begin{eqnarray}
\lefteqn{\Gamma_{Z\rightarrow \mbox{\scriptsize hadrons}}
        =\frac{G_FM_Z^3}{8\sqrt{2}\pi}}
\nonumber\\
&&\times \Biggl\{\sum_f  \biggl(
\rho_f v_f^2\biggl[(1+2X_f)\sqrt{1-4X_f}
+\delta_{\mbox{\tiny QCD}}^{\mbox{\tiny V}}(\alpha_s,X_f,X_t)
+\delta_{\mbox{\tiny QED}}^{\mbox{\tiny V}}(\alpha,\alpha_s,X_f)
                                     \biggr]
\nonumber\\
 &&  \hspace{15mm}
+\rho_f a_f^2\biggl[(1-4X_f)^{3/2}
+\delta_{\mbox{\tiny QCD}}^{\mbox{\tiny A}}(\alpha_s,X_f,X_t)
+\delta_{\mbox{\tiny QED}}^{\mbox{\tiny A}}(\alpha,\alpha_s,X_f)
                                     \biggr]
                                     \biggr)
\nonumber\\
&&\quad +\ {\cal L}^{\mbox{\scriptsize V}}
+{\cal L}^{\mbox{\scriptsize A}}
\Biggr\}.
\label{Zqq}
\end{eqnarray}
Here there is a sum over light quark flavors $f = u,d,s,c,b$. We
define $X_f=m_f(M_Z)^2/M_Z^2$ and $X_t=m_t(M_Z)^2/M_Z^2$. (We
use the $\overline{\rm MS}$ definition of masses.)

The vector and axial couplings of quark $f$ to the $Z$ boson are
$v_f=\{2I_f^{(3)}-4e_f\sin^2\theta_{\mbox{\tiny W}}k_f\}$ and
$a_f=\,2I_f^{(3)}$. The electroweak self-energy and vertex
corrections are absorbed in the factors $\rho_f$ and $k_f$. The
current status of the  electroweak contributions has been discussed
in detail in Ref.~\cite{revZ}. The small QED corrections in vector and
axial channels have the form
\begin{equation}
\delta_{\mbox{\tiny QED}}^{\mbox{\tiny V}}
     =\frac{3}{4}\,e_f^2\,\frac{\alpha}{\pi}\,[1+12X_f+O(X_f^2)]
                       +O(\alpha^2)+O(\alpha\alpha_s),
\label{QEDV}
\end{equation}
\begin{equation}
\delta_{\mbox{\tiny QED}}^{\mbox{\tiny A}}
     =\frac{3}{4}\,e_f^2\,\frac{\alpha}{\pi}\,[1-6X_f
                   -12X_f\log X_f+O(X_f^2)]
                       +O(\alpha^2)+O(\alpha\alpha_s).
\label{QEDA}
\end{equation}
The corrections of order $\alpha^2$ and $\alpha\alpha_s$ are
discussed in Ref.~\cite{katQED}.

It is convenient to decompose the QCD contributions into singlet and
non-singlet parts and further into vector ($V$) and axial vector
($A$) contributions. The nonsinglet parts are  represented by the
terms $\delta_{\mbox{\tiny QCD}}^{\mbox{\tiny V}}$ and
$\delta_{\mbox{\tiny QCD}}^{\mbox{\tiny A}}$, and correspond to
cut Feynman graphs in which a single quark loop of flavor $f$
connects the two electroweak current operators. The singlet
contributions correspond to graphs with the electroweak currents in
separate quark loops mediated by gluonic states. In the singlet
contributions one does not have a single sum over a flavor $f$. These
contributions are represented by the terms ${\cal L}^{\mbox{\scriptsize
V}}$ and ${\cal L}^{\mbox{\scriptsize A}}$.

The nonsinglet QCD contribution in the vector channel to order
$\alpha_s^3$ can be written in the form
\begin{eqnarray}
\delta_{\mbox{\tiny QCD}}^{\mbox{\tiny V,\scriptsize ns}}
&=&\frac{\alpha_s}{\pi}\,\left[1+12\,{X}_f\right]
\nonumber\\
&&+\biggl(\frac{\alpha_s}{\pi}\biggr)^2
\left[1.40923 + 104.833\,{X}_f
+\sum_v F^{(2)}(X_v) + G^{(2)}(X_t)\right]
\nonumber\\
&&
+\biggl(\frac{\alpha_s}{\pi}\biggr)^3
\left[-12.76706 + 547.879\,{X}_f
     +\sum_v F^{(3)}(X_v) + G^{(3)}(X_t)\right].
\label{Vns}
\end{eqnarray}
In this formula, $\alpha_s$ denotes the running $\overline{\rm MS}$
coupling in five flavor theory evaluated at $M_Z$. The transformation
relation for different number of flavors and different scales, as
well as the relation between the $\overline{\rm MS}$ running mass and
the pole mass can be found in Ref.\ \cite{H}.

The order $\alpha_s^2$ and $\alpha_s^3$ terms have been evaluated in
the limit of vanishing light quark masses and infinitely large top
mass in Refs.~\cite{R2,R3}. These contributions, $(\alpha_s/\pi) +
1.40923\,(\alpha_s/\pi)^2 -12.76706\,(\alpha_s/\pi)^3$, are the
\{vector,nonsinglet\} part of the perturbative series analyzed
in the main body of this paper.

The terms proportional to ${X}_f$ represent the leading corrections
to the approximation $X_f= 0$, as given in Ref.~\cite{O(m)}.

The function $F^{(2)}(X_v)$  arises from three-loop diagrams
containing an internal quark loop with a  quark of flavor $v =
u,d,s,c,b$ propagating in it (while the quark of flavor
$f$ couples to the weak currents). This function represents the
corrections to the approximation $X_v= 0$. These contributions are
already small, so it suffices to approximate $X_f$ by 0 in
$F^{(2)}(X_v)$.  In fact, numerically \cite{Mt},
\begin{equation}
F^{(2)}(X_v) \approx {X}_v^2 \times \left\{
-0.474894
- \log {X}_v
+\sqrt{{X}_v}  \left[
-0.5324
+0.0185 \log {X}_v \right]
\right\}
\end{equation}
is so small that the whole function could be neglected.

The function  $G^{(2)}(X_t)$ represents the contribution of virtual
top quark loops inside three-loop cut Feynman diagrams. These
contributions are small since the top quark is nearly decoupled from
the theory. Thus it suffices to approximate $X_f$ by 0 in
$G^{(2)}(X_t)$. Numerically, one finds \cite{Mt}
\begin{equation}
G^{(2)}(X_t) \approx { {X}_t^{-1}}\times\left\{
  {44 \over 675}
+ {2 \over 135} \log {X}_t
-\sqrt{{X}_t^{-1}}
\left[0.001226
+0.001129\log {X}_t
\right]
\right\}.
\label{Glargem}
\end{equation}
The first two terms in the right hand side of eq.(\ref{Glargem}) have
also been obtained using the large mass expansion
method \cite{Chetvirt}.

At order $\alpha_s^3$, there can be two internal quark loops.
However, it suffices to consider only one loop with a nonzero light
quark mass at a time, or one top quark loop with all light quark
masses set to zero. Then we can define functions $F^{(3)}(X_v)$ and
$G^{(3)}(X_t)$ analogously to $F^{(2)}(X_v)$ and $G^{(2)}(X_t)$.
For $F^{(3)}(X_v)$ the following small mass expansion
is obtained in Ref.~\cite{O(m)}:
\begin{equation}
F^{(3)}(X_v) \approx -6.12623 {X}_f.
\end{equation}
For $G^{(3)}(X_t)$ the following large mass expansion has been
obtained in Ref~\cite{Lar}:
\begin{equation}
G^{(3)}(X_t) \approx { {X}_t^{-1}}\times [
-0.1737-0.2124\log {X}_t - 0.0372 \log^2{X}_t].
\label{Glargem3}
\end{equation}

The nonsinglet contribution in the axial channel is the same as
the one in the vector channel except that the contributions
proportional to $X_f$ \cite{O(m),Mt} are different:
\begin{eqnarray}
\delta_{\mbox{\tiny QCD}}^{\mbox{\tiny A,\scriptsize ns}}
&=&\frac{\alpha_s}{\pi}\,\left[1-22\,{X}_f\right]
\nonumber\\
&&+\biggl(\frac{\alpha_s}{\pi}\biggr)^2
\left[1.40923 - 85.7136\,{X}_f
+\sum_v F^{(2)}(X_v) + G^{(2)}(X_t)\right]
\nonumber\\
&&+\biggl(\frac{\alpha_s}{\pi}\biggr)^3
\left[-12.76706 + (\mbox{\small unknown})\,{X}_f
+\sum_v F^{(3)}(X_v) + G^{(3)}(X_t)\right].
\label{Ans}
\end{eqnarray}

We now turn to the singlet contributions, which start at order
$\alpha_s^2$:
\begin{equation}
{\cal L}^{\mbox{\scriptsize V/A}} =
{\cal L}_2^{\mbox{\scriptsize V/A}} \left(\alpha_s \over \pi \right)^2
+ {\cal L}_3^{\mbox{\scriptsize V/A}} \left(\alpha_s \over \pi \right)^3
+\cdots.
\end{equation}
At order $\alpha_s^2$, there is no vector contribution,
\begin{equation}
{\cal L}_2^{\mbox{\scriptsize V}}=0,
\end{equation}
while the axial contributions from $u$ and $d$ quarks and from $c$
and $s$ quarks vanish in the limit of vanishing quark masses. This is
because in the Standard Model the quarks in a weak doublet couple
with the opposite sign to the weak axial current. However, the
contribution from the $t$,$b$ doublet is significant because of the
large mass splitting \cite{tb}:
\begin{eqnarray}
{\cal L}_2^{\mbox{\scriptsize A}}&=&
-\frac{37}{12}-\log X_t+\frac{7}{81}X_t^{-1}
+0.013X_t^{-2}
\nonumber\\ &&
+{X}_b\,(18 +6\,\log X_t)
-\frac{{X}_b}{X_t}\,\biggl(\frac{80}{81}+\frac{5}{27}\log X_t\biggr).
\label{Axtri}
\end{eqnarray}
Here the corrections proportional to ${X}_b$ have been calculated in
Ref.~\cite{Chettri}.

At order $\alpha_s^3$, both channels contribute.
The vector contribution in the limit of massless light
quarks is \cite{R3}
\begin{equation}
{\cal L}_3^{\mbox{\scriptsize V}}=
-0.41318\,\biggl( \sum_{f} v_f \biggr)^2
+\left(0.02703\,{X}_t^{-1}+0.00364\,{X}_t^{-2}+O(X_t^{-3})\right)\,
v_t\sum_{f}v_f.
\label{Vtri3}
\end{equation}
The sums here run over light quark flavors $f = u,d,s,c,b$. The terms
proportional ${X}_t^{-1}, {X}_t^{-2}$ were computed in
Ref.~\cite{Lar} and turn out to be negligible.

In the axial channel, the order $\alpha_s^3$ singlet contribution
in the large top mass expansion reads \cite{Chettri3,Lar,revZ}
\begin{equation}
{\cal L}_3^{\mbox{\scriptsize A}}=
-15.98773 -\frac{67}{18}\log X_t + \frac{23}{12}\log ^2X_t.
\label{Atri3}
\end{equation}
Corrections for a nonzero $b$ quark mass are not yet known. However,
at the level of precision of this paper, they are not expected to be
significant.

\end{document}